\documentstyle[emulateapj,epsf]{article}

\def\rms{\hbox{\it rms\/}}

\def\fivesecavgs{\hbox{5s averages}}
\def\Ka{$K_a$}
\def\Q{$Q$}
\def\D{$D$}
\def\tilde{\char'176}

\slugcomment{Submitted to ApJL 1999 April 30; revised 1999 June 10}

\lefthead{Torbet et al.}
\righthead{TOCO97}

\begin{document}

\title{A Measurement of the Angular Power Spectrum of the Microwave
Background Made from the High Chilean Andes}

\author{E. Torbet\altaffilmark{1}, M. J. Devlin\altaffilmark{2},
W. B. Dorwart\altaffilmark{1}, T. Herbig\altaffilmark{1}, 
A. D. Miller\altaffilmark{1}, M. R. Nolta\altaffilmark{1}, 
L. Page\altaffilmark{1}, J. Puchalla\altaffilmark{2}, 
H. T. Tran\altaffilmark{1}}

\altaffiltext{1}{Princeton University, Physics Department, Jadwin Hall,
Princeton, NJ 08544} 
\altaffiltext{2}{University of Pennsylvania, Department of Physics and
Astronomy, David Rittenhouse Laboratory, Philadelphia, PA 19104} 

\begin{abstract}

We report on a measurement of the angular spectrum of the anisotropy
of the microwave sky at 30 and 40~GHz between $l=50$ and $l=200$.
The data, covering roughly $600\rm\,deg^2$, support a rise in the
angular spectrum to a maximum with $\delta T_l \approx 85~\mu$K at $l=200$.
We also give a 2-sigma upper limit of $\delta T_l < 122~\mu$K
at~$l=432$ at 144 GHz.
These results come from the first campaign of the
Mobile Anisotropy Telescope ({\sl MAT}) on Cerro Toco, Chile.
To assist in assessing the site, we present plots of the
fluctuations in atmospheric emission at 30 and 144 GHz.

\end{abstract}


\section{Introduction}

The characterization of the CMB anisotropy is essential for
understanding the process of cosmic structure formation (e.g. \cite{whu97}).
If some of the currently popular models prove correct,
the anisotropy may be used to strongly constrain
cosmological parameters (\cite{jung95}, \cite{bond98}).
Here we report the results from the {\sl TOCO97} campaign of the Mobile
Anisotropy Telescope ({\sl MAT}) experiment. 

%
%

\section{Instrument}
\label{inst}

The {\sl MAT} telescope is comprised of the {\sl QMAP} balloon
gondola and instrument (\cite{dev98}), mounted on the azimuthal
bearing of a surplus Nike Ajax military radar
trailer\footnote{Details of the experiment, 
synthesis vectors, data, and analysis code may be found at
http://www.hep.upenn.edu/CBR/ and
http://pupgg.princeton.edu/\tilde{}cmb}.
The receiver has five cooled corrugated feed horns,
one at \Ka~band (31\,GHz), two at \Q~band (42\,GHz),
and two at \D~band (144\,GHz).
Each of the \Ka\ and \Q~band horns feed two HEMT-based
(high electron mobility transistor) amplifiers
(\cite{posp92}) with one in each polarization.
The two D~band horns each feed a single SIS detector (\cite{ker93})
with one horn in each polarization.
This gives a total of eight radiometry channels in the
experiment\footnote{HEMT amplifiers have improved considerably
since this time (\cite{posp97}) and SIS receivers are generally
more sensitive than what we achieved. In 1997, one of the \D\ channels
and one of the \Q\ channels did not have sufficient sensitivity
to warrant a full analysis.}.
A Sumitomo mechanical refrigerator cools the HEMT amplifiers to 35~K
and the SIS receivers to 4~K. 


The telescope optics are similar to those used for three
ground-based observing campaigns in Saskatoon, SK (\cite{wol97}, {\sl SK}).
The feeds underilluminate an ambient temperature 0.85\,m off-axis
parabolic reflector which in turn underilluminates a  computer
controlled 1.8\,m$\times$1.2\,m resonant chopping flat mirror.
The beams are scanned horizontally across the
sky in a $\approx 4.6$~Hz sinusoidal pattern.
The outputs of the detectors are AC coupled at 0.15~Hz
and sampled $N_c$ times during each chopper cycle
($N_c = 80$ for \Ka\ and \Q~bands, and $N_c = 320$ for \D~band).
The telescope is inside an aluminum ground screen which is fixed with
respect to the receiver and parabola.  

%
%
%
%

%
%
%

The telescope pointing (Table~1) is established through
observations of Jupiter and is monitored with two redundant encoders on
both the azimuth bearing and on the chopper.
The absolute errors in azimuth and elevation are 0\fdg04,
and the relative errors are $<$0\fdg01.
The chopper position is sampled 80 times per chop.
When its {\it rms\/} position over one cycle deviates by more than
0\fdg015 from the average position (due to wind, etc.),
we reject the data.

%
%

\smallskip
\begin{center}
\vbox{Table 1. TOCO97 beam characteristics}
\smallskip
\begin{tabular}{ccccc}
\tableline\tableline
\noalign{\vskip .4em}
Feed & Az & El & $\Omega_{meas}$ & $\theta^{FWHM}_{avg}$ \\
 & deg & deg & $10^{-4}~$sr & deg \\
\noalign{\vskip .4em}
\tableline
\noalign{\vskip .4em}
\Ka1/2 & 203.13 & 41.75 & 2.75 & 0.90 \\
\Q1/2  & 206.75 & 41.85 & 1.69 & 0.70 \\
\Q3/4  & 206.70 & 39.25 & 1.77 & 0.72 \\
\D1    & 205.00 & 40.44 & 0.183 & 0.23 \\
\noalign{\vskip .4em}
\tableline
\end{tabular}
\end{center}

\section{Observations and Calibration}

\begin{figure*}
\epsscale{2}
\plotone{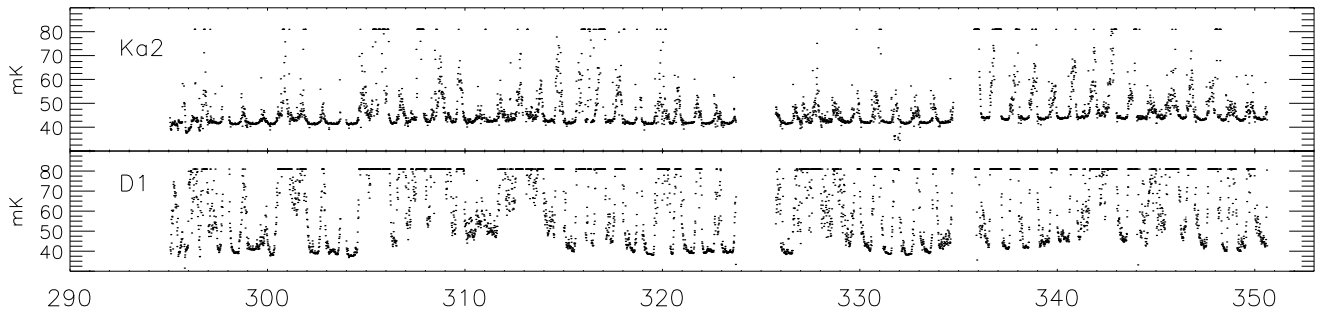}
\caption{The {\it rms} detector output in
antenna temperature of the \Ka2 and \D1\ channels
averaged over 0.68 ms (with the chopper running at the nominal
amplitude) vs. day of year in 1997.
The sky is most stable between 10 PM and 10 AM local time. Similar
results from 1998 are consistent with the NRAO opacity
measurements (http://www.tuc.nrao.edu/mma/sites/sites.html).
\label{fig-rms}}
\end{figure*}

Data were taken at a 5200 m site\footnote{ The Cerro Toco site of
the Universidad Catolica de Chile was made available through the
generosity of Prof. Hern\'{a}n Quintana, Dept. of Astronomy and
Astrophysics. It is near the proposed MMA site.} on the side of
Cerro Toco (lat. = -22\fdg95 long. = 67\fdg775 ), near San Pedro de
Atacama, Chile, from Oct. 20, 1997 to Dec 15, 1997. The receiver
was operational 90\% of the available time. For the anisotropy
data, the primary optical axis is fixed at {\it az} = 204\fdg9,
{\it el} = 40\fdg5, $\delta$ = -62\fdg6 and the
chopper scans with an azimuthal amplitude of $2\fdg96$
(8\fdg93 on the sky) as the sky rotates through the beam.  The
telescope position was not wobbled to the other side of the South
Celestial Pole as for the {\sl SK} measurements in
the North.  The {\it rms} outputs  of the \Ka2\ and \D1 channels
are shown in Fig.~1. 

Jupiter is used to calibrate all channels and map the beams.
Its brightness temperature is 152, 160, 170~K for \Ka\ through \D~bands
respectively (\cite{gri86}, \cite{uli81}), with an intrinsic
calibration error of $\approx 5\%$.
We account for the variation in angular diameter.
We also observe Jupiter with multiple relative azimuthal offsets
to verify the chopper calibration.

The uncertainty in the beam solid angle for the \Ka\ and \Q\ bands is
$\approx 5\%$ as determined from the standard deviation of beam measurements
for the {\sl MAT} and {\sl QMAP} experiments.
From a global fit of the clear-weather Jupiter calibrations,
the standard deviation in the fitted amplitudes is 6\%.
These sources of calibration error dominate the error from the uncertainty in
the passband.
The total $1\sigma$ calibration error is obtained by
combining the intrinsic, beam, and measurement errors in
quadrature resulting in 10\%, 10\%, and 11\% in \Ka\ through \D\ respectively.

A thermally-stabilized noise source at $T_{\rm eff}\approx 1$~K is
switched on twice for 40~msec every 100 seconds as a 
relative calibration. The pulse height is correlated to the
Jupiter calibrations in the \Ka\ and \Q\ channels. The variation in
detector gain corrected for with these calibration pulses is roughly 5\%. 
No such correction was made for \D~band.

\section{Data Reduction}

The reduction is similar to that of the SK experiment
(\cite{net97}). The raw data, $d_i$, are multiplied by ``n-pt''
synthesis vectors, $SV_{n,i}$ (where $i$ ranges from 1 to $N_c$)
to yield the effective temperature corresponding to a  multilobed
beam on the sky, $H(\Omega)$. For example, we refer to the
classic three-lobed beam produced by a ``double difference'' as
the ``3-pt harmonic'' and write $t_3=\sum_{i=1}^{N_c} SV_{3,i}
d_i$. We also generate the quadrature signal $q_n$ (data with
chopper sweeping in one direction minus that with the sweeping in
the other direction) and fast-dither signal $f^d_n$ (one value of
$t_n$ minus the subsequent one). For both \Ka1/2 and \Q3/4 we
analyze the unpolarized weighted mean of the combined detector
outputs.

The phase of the data relative to the beam position is determined
with both Jupiter and observations of the galaxy. We know we are
properly phased when the quadrature signal from the galaxy is
zero for all harmonics.

%
%

The harmonics are binned according to the right ascension at the
center of the chopper sweep. The number of bins
depends on the band and harmonic (Table~2). For each 
night, we compute the mean and variance of all the $t_n$,
$q_n$, and $f^d_n$ corresponding to a bin. These numbers  are
appropriately averaged over the campaign and  used in the
likelihood analysis.

From the raw dataset of 814250~\fivesecavgs, we filter out time
spent on instrument calibration (6\%), celestial calibrations
(11\%), observations of the galaxy \& daytime (53\%), and bad pointing
(4\%). Accounting for overlap, these cut a total of 57\%. The
data span RA = $0^{\circ}$ to $140^{\circ}$ ($b=-55^{\circ}$ to
$-10^{\circ}$).  


%
%

The data are selected according to the weather by examining each
harmonic independently. We first flag \fivesecavgs\ with a large
\rms. The unflagged data are divided up into 15~minute sections
and the \rms\ of the $t_n$ found. For 15~m sections with 
\rms\ $>2\sigma$, the constituent \fivesecavgs\ are not used, as well as those
of the preceding and succeeding 15~m sections.
We ensure that the cut does not bias the statistical 
weight. As a final cut, nights with less than 4.7 hours of data are
excluded. Repeating the analysis with increased cut values
produces statistically similar (within 1$\sigma$) results. The atmosphere
cut selects roughly the same sections for \Ka\ and \Q. In the
analyses, we discard the 2 and 3-pt data as it is corrupted by
atmospheric fluctuations and variable instrumental offsets. If
the 4-pt is corrupted, it is at the $1\sigma$ level and not
readily detectable.

%
%
%
%
%
%

The stability of the instrument is assessed through internal
consistency checks and with the distribution
of the offset of each harmonic. The offset is the average of a
night of data after the cuts have been applied (ranges
from 5-10~hours) and is of magnitude $\approx 200~\mu$K with
error $20~\mu$K. 
In general, the offset remains constant for a few
nights and then jumps 3-5 sigma. The resulting $\chi^2/\nu$ is typically
between 4 and 20 for the data over the full campaign and is $\approx 1$ for
the quadrature signal. In general, a change in offset can have
any time scale. The $q_n$ and $f^d_n$ are sensitive to
$\tau=0.25~$s. We also monitor a slow dither (difference of the
subsequent 5~sec averages) with $\tau=5$~s and a night-to-night dither
with $\tau=24~$h. For the final
analysis, we delete one seven day section that has a large jump in
offset. To eliminate the potential effect of slow variations in
offset, we remove the slope and mean for each night. This is
accounted for in the quoted result (both the constraint matrix
method, \cite{bond98b}, and marginalization, \cite{bond91} give
similar corrections) and does not significantly alter the results
over the subtraction of a simple mean. As a test, we have also
tried removing quadratic and cubic terms from the offset, with no significant changes in the
answer.
In summary, there is no evidence that the small instability in the
offset affects our results.


We examined the variations in the power spectrum of the synchronously
co-added raw HEMT data, and found no evidence for microphonics.
However, a microphonic coupling to the SIS detector
was exacerbated after situating the telescope at the site. After
filtering, residual signals persisted in the quadrature channels
(though not in the fast and slow dithers) and so we report only
95\% upper limits for the D channel, specifically  $\delta T_l < 180~\mu$K
at $\ell = 325$ and $\delta T_l < 122~\mu$K at $\ell = 432$.

The primary effect of data editing is to increase the error bar per point
and decrease the upper limits of the null tests.
Of the 169 null tests (Table~2
plus fast, slow, and night dithers),
there are only three failures.
The distribution of the reduced $\chi^2$ of the null tests is
consistent with noise and inconsistent with any signal.
When the data are combined into groups of harmonics and bands,
all null tests are consistent with noise.

\section{Analysis and Discussion}

The analysis of the individual harmonics, because the windows are
so narrow, essentially corresponds to finding $\delta T_l^\prime
= \sqrt{(\Delta^2_{tot} - \Delta^2_{inst})/I(W)}$, where
$\Delta^2_{tot}$ is the variance of the data for each harmonic,
$\Delta^2_{inst}$ is the variance due to atmospheric and
instrumental noise, and $I(W)=\sum W_l/l$. $W_l$ is the
window function, as defined in \cite{bond96}. The full likelihood
analysis provides a formal way of determining $\delta T_l$ that
includes correlations and gives the correct error bar in the low
signal-to-noise limit.

The error in $I(W)$ is determined from the scatter in the beam
values. We find $\delta I(W)/I(W) \lesssim 0.01$ for all bands and harmonics.
The mean variance,
$\Delta^2_{inst}$, is determined directly from the uncertainties
in each bin. If these uncertainties are somehow biased, the
results of the simple test and full likelihood will be biased. We
examine the distribution of all the data for each harmonic from
all the nights after removing the mean value of each sky bin. The
width of this distribution agrees with the mean error bar
indicating that the error per point is not biased. Also, the
ratio of the error bars between harmonics agrees with the
analytic calculation.

In the full analysis (Fig.~2),
we include all known correlations inherent
in the observing strategy. From the data, we determine the
correlations between harmonics due to the atmosphere, detector
noise, and non-orthogonality of the synthesis vectors.
The correlation coefficients between bands due to the atmosphere
are of order~$0.05$. We also examine the autocorrelation function
of the data for a single harmonic to ensure that atmospheric
fluctuations do not correlate one bin to the next. The quoted
results are insensitive to the precise values of the off-diagonal
terms of the covariance matrix.

%
%
%
%

\medskip
\centerline{\vbox{\epsfxsize=5cm\epsfbox{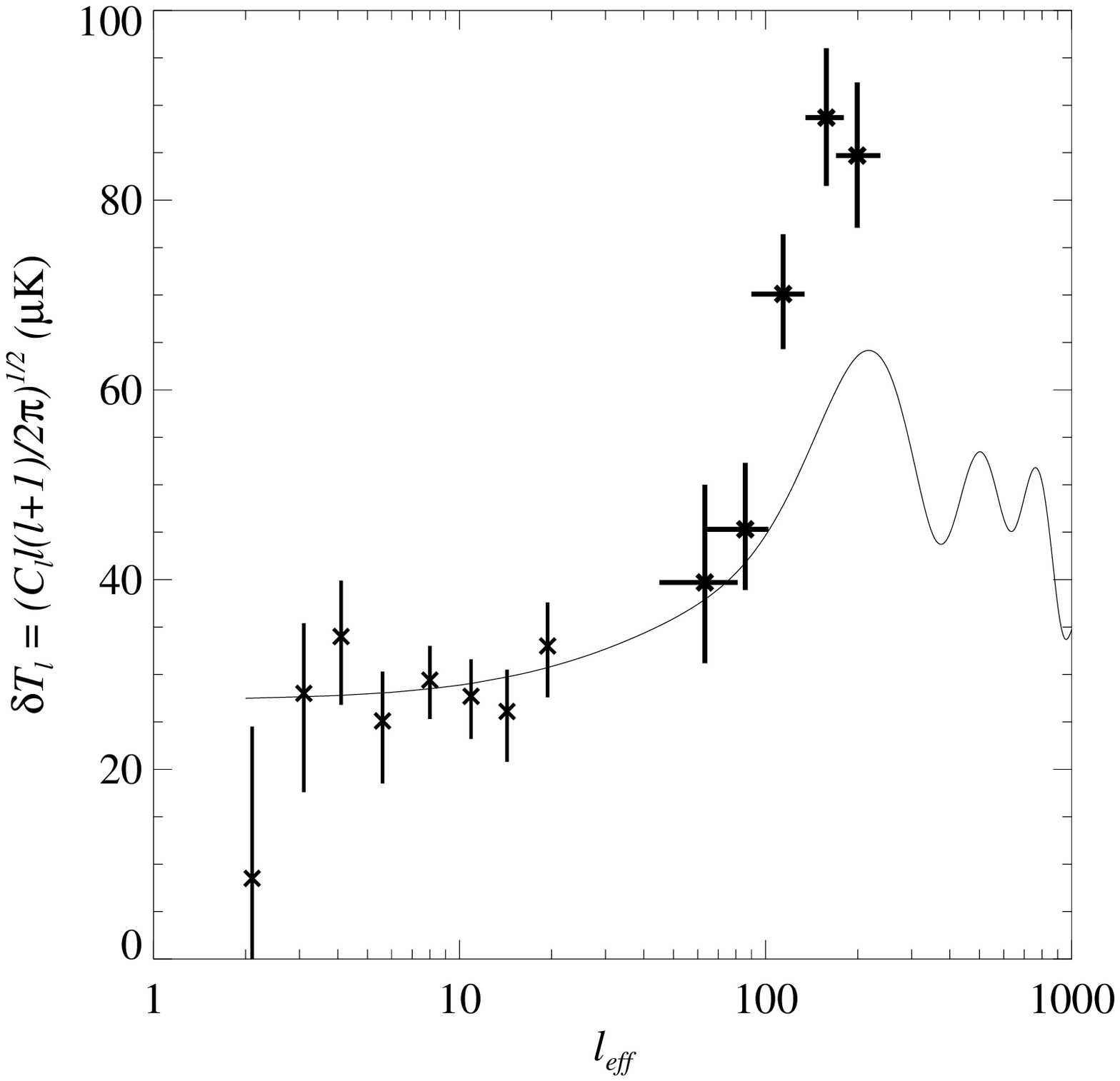}}}
{\small
F{\scriptsize IG}.~2.---
Combined analysis of data in Table~2. The values are
($l,\delta T_l~[\mu{\rm K}]$) $=$
($63^{+18}_{-18}, 40^{+10}_{-9}$), ($86^{+16}_{-22}, 45^{+7}_{-6}$),
($114^{+20}_{-24}, 70^{+6}_{-6}$), ($158^{+22}_{-23}, 89^{+7}_{-7}$),
($199^{+38}_{-29}, 85^{+8}_{-8}$).  Error bars are ``$1\sigma$ statistical'';
calibration error
is not included. The {\sl COBE}/DMR points are from \cite{max97}.
The solid curve is standard CDM ($\Omega_b=0.05$, $h=0.5$).
\label{fig-pspec}}
\medskip

These results are similar to previous results obtained  with this
technique (SK) though the experiment was done  with different optics,
a different receiver, a different primary calibrator, largely
different analysis code, and observed a different part of the
sky. Though we have not correlated our data with templates of
foreground emission, the foreground contribution is known to be
small at these frequencies and galactic latitudes (\cite{cob99},
\cite{doc97}).  In addition we have examined the frequency spectrum of
the fluctuations in \Ka\ and \Q~bands, and find it to be consistent
with a thermal CMB spectrum, and inconsistent with various foregrounds.
Finally, the full analysis has been
repeated after deleting each 15\arcdeg\ section of data in RA, indicating
that the signal does not arise from one region. (Our scan
passes near, but misses, the LMC.) Future work will address
the precise level of contamination.

%
%

\acknowledgments

We gratefully acknowledge the insights and help from Dave
Wilkinson, Norm Jarosik, Ray Blundell, Angel Ot\'{a}rola, Hern\'{a}n
Quintana, Robert Caldwell, Ted Griffith, Bernard Jones, and
Harvey Chapman. The experiment would not have been possible
without NRAO's site monitoring and detector development.  We also
thank Lucent Technologies for donating the radar trailer.  This
work was supported by an NSF NYI award, a Cottrell Award from the
Research Corporation, a David and Lucile Packard Fellowship (to
LP), a NASA GSRP fellowship to AM, an NSF graduate fellowship to MN, 
NSF grants PHY-9222952, PHY-9600015, AST-9732960, and the
University of Pennsylvania.  The data will
be made public upon publication of this {\it Letter}.

\begin{deluxetable}{rccccccccccc}
\tablenum{2}
\footnotesize
\tablecaption{TOCO97 Angular Spectrum \label{tbl-pspec}}
\tablewidth{18cm}
\tablehead{
\colhead{Band/$SV$} &
\colhead{$\ell_{\hbox{eff}}$\,\tablenotemark{a}} &
\colhead{$\delta T_\ell$\,\tablenotemark{b}} &
\colhead{$N_{bins}$\,\tablenotemark{c}} &
\colhead{$\delta T_\ell^\prime$} &
\colhead{$\Delta_{tot}$} &
\colhead{$\Delta_{inst}$} &
\colhead{$\sqrt{I(W)}$} &
\colhead{(A-B)/2\,\tablenotemark{d,e}} &
\colhead{Quad, q$_n$\,\tablenotemark{d}} &
\colhead{(H1-H2)/2\,\tablenotemark{d,f}} \nl
 & & $\mu$K &  & $\mu$K & $\mu$K & $\mu$K & &
$\mu$K & $\mu$K & $\mu$K }
\startdata
\sidehead{\Ka1/2}
4pt&$63^{+17}_{-18}$&$35^{+13}_{-9}$&48(16)&32&33&20&0.84&$<27 (0.94)$&$<30 (1.05)$&$<29 (0.95)$\nl
5pt&$86^{+16}_{-21}$&$52^{+11}_{-8}$&64(28)&49&40&21&0.71&$<21 (0.59)$&$<32 (1.23)$&$<29 (1.05)$\nl
6pt&$107^{+16}_{-21}$&$71^{+12}_{-10}$&96(42)&69&52&27&0.65&$<32 (0.90)$&$<31 (0.96)$&$<35 (1.04)$\nl
7pt&$127^{+16}_{-22}$&$93^{+15}_{-14}$&96(41)&90&57&27&0.55&$<35 (0.80)$&$<37 (0.90)$&$<30 (0.68)$\nl
8pt&$145^{+18}_{-20}$&$103^{+15}_{-13}$&128(55)&102&63&34&0.52&$<51 (0.97)$&$<46 (1.00)$&$<43 (0.91)$\nl
9pt&$165^{+18}_{-20}$&$65^{+16}_{-17}$&128(54)&59&47&38&0.46&$<63 (0.95)$&$<72 (1.19)$&$<66 (1.07)$\nl
10pt&$182^{+21}_{-17}$&$67^{+20}_{-23}$&192(82)&70&60&51&0.44&$<68 (0.85)$&$<69 (0.94)$&$<60 (0.89)$\nl
11pt&$192^{+30}_{-8}$&$<119$&192(82)&67&65&58&0.42&$<91 (0.94)$&$<83 (1.00)$&$<86 (0.96)$\nl
12pt&$215^{+27}_{-11}$&$128^{+30}_{-33}$&192(82)&127&83&68&0.37&$<150 (1.11)$&$<86 (0.79)$&$<76 (0.67)$\nl
\sidehead{\Q1}
4pt&$63^{+17}_{-18}$&$57^{+18}_{-13}$&48(20)&51&53&31&0.83&\nodata&$<44 (1.04)$&$<47 (1.11)$\nl
5pt&$87^{+16}_{-22}$&$40^{+14}_{-14}$&64(28)&34&40&33&0.71&\nodata&$<36 (0.75)$&$<47 (1.09)$\nl
6pt&$110^{+15}_{-24}$&$56^{+14}_{-13}$&96(42)&52&52&40&0.65&\nodata&$<45 (0.87)$&$52^{+15}_{-14} (1.71)$\nl
7pt&$131^{+14}_{-25}$&$81^{+19}_{-16}$&96(42)&77&59&41&0.55&\nodata&$<65 (1.12)$&$<53 (0.89)$\nl
8pt&$151^{+15}_{-25}$&$86^{+19}_{-17}$&128(55)&79&66&50&0.53&\nodata&$<45 (0.67)$&$<64 (0.89)$\nl
9pt&$172^{+14}_{-26}$&$93^{+23}_{-23}$&128(55)&89&68&54&0.47&\nodata&$<82 (0.94)$&$<72 (0.83)$\nl
10pt&$191^{+16}_{-24}$&$<115$&192(84)&31&75&74&0.47&\nodata&$<105 (1.13)$&$<92 (0.94)$\nl
11pt&$203^{+24}_{-17}$&$<117$&192(84)&44&80&78&0.44&\nodata&$<103 (0.94)$&$<83 (0.81)$\nl
12pt&$221^{+25}_{-15}$&$<169$&192(84)&91&91&84&0.40&\nodata&$<138 (1.06)$&$<119 (0.92)$\nl
13pt&$245^{+21}_{-20}$&$<130$&192(84)&\nodata&84&92&0.37&\nodata&$<163 (1.02)$&$<164 (1.06)$\nl
14pt&$267^{+20}_{-23}$&$<202$&256(112)&123&123&114&0.37&\nodata&$<183 (1.06)$&$<188 (1.00)$\nl
\sidehead{\Q3/4}
5pt&$83^{+15}_{-20}$&$47^{+17}_{-13}$&64(17)&43&39&24&0.73&$31^{+10}_{-8} (1.94)$&$<50 (1.39)$&$<58 (1.73)$\nl
6pt&$106^{+14}_{-23}$&$61^{+18}_{-13}$&96(22)&55&46&28&0.66&$<27 (0.68)$&$<34 (0.60)$&$<61 (1.46)$\nl
7pt&$125^{+14}_{-23}$&$72^{+16}_{-12}$&96(35)&71&50&30&0.56&$<29 (0.65)$&$<57 (1.30)$&$<31 (0.57)$\nl
8pt&$145^{+14}_{-23}$&$115^{+19}_{-15}$&128(45)&109&68&35&0.54&$<26 (0.50)$&$<38 (0.73)$&$<60 (1.18)$\nl
9pt&$165^{+14}_{-24}$&$72^{+24}_{-21}$&128(29)&65&48&37&0.48&$<43 (0.82)$&$<97 (1.38)$&$<84 (1.16)$\nl
10pt&$184^{+15}_{-23}$&$87^{+19}_{-19}$&192(70)&86&63&48&0.47&$<68 (1.03)$&$<45 (0.78)$&$<52 (0.75)$\nl
11pt&$196^{+22}_{-17}$&$90^{+27}_{-26}$&192(54)&84&65&53&0.44&$<64 (0.90)$&$<78 (0.91)$&$<75 (0.78)$\nl
12pt&$212^{+25}_{-13}$&$100^{+29}_{-27}$&192(65)&97&69&57&0.40&$<70 (0.95)$&$<129 (1.36)$&$<84 (0.89)$\nl
13pt&$236^{+20}_{-19}$&$<157$&192(56)&80&71&65&0.37&$<107 (1.07)$&$<121 (0.93)$&$<153 (1.17)$\nl
14pt&$258^{+18}_{-23}$&$119^{+36}_{-38}$&256(103)&119&91&80&0.36&$<116 (0.95)$&$<137 (1.11)$&$<103 (0.78)$\nl
\enddata
\tablecomments{A ``$<$'' indicates a 95\% confidence limit.
Calibration errors are {\bf not} included.
(a) The range for $\ell_{\hbox{eff}}$ denotes
the range for which the window function exceeds $e^{-1/2}$ times
the peak value.
(b) The error on $\delta T_\ell=[\ell(\ell+1)C_\ell/2\pi]^{1/2}$ is
comprised of experimental uncertainty and sample variance.
These values are not statistically independent: harmonic numbers
differing by 2 are correlated at the 0.35 level.
For all harmonics, the sample variance ($\propto 1/\sqrt{2N_{bins}}$)
is $\approx 7~\mu$K.
(c) The number of bins on the sky followed by, in parentheses,
the number used in the analysis due to the galactic/atmosphere cut.
(d) The reduced $\chi^2$ are given in parentheses.
(e) $(A-B)/2$ is the difference in polarizations.
We have combined bands and harmonics to generate
95\% upper limits on polarization, $A-B$, and obtain: 
($l,\delta T_l~[\mu{\rm K}]$) $=$
($63^{+18}_{-18}, <37$), ($86^{+17}_{-21}, <54$),
($115^{+21}_{-24}, <28$), ($148^{+17}_{-25}, <41$),
($195^{+33}_{-23}, <79$).
(f) $(H1-H2)/2$ is the first half minus the second half.}
\end{deluxetable}


\newpage

\begin{figure}
\epsscale{1}
\plotone{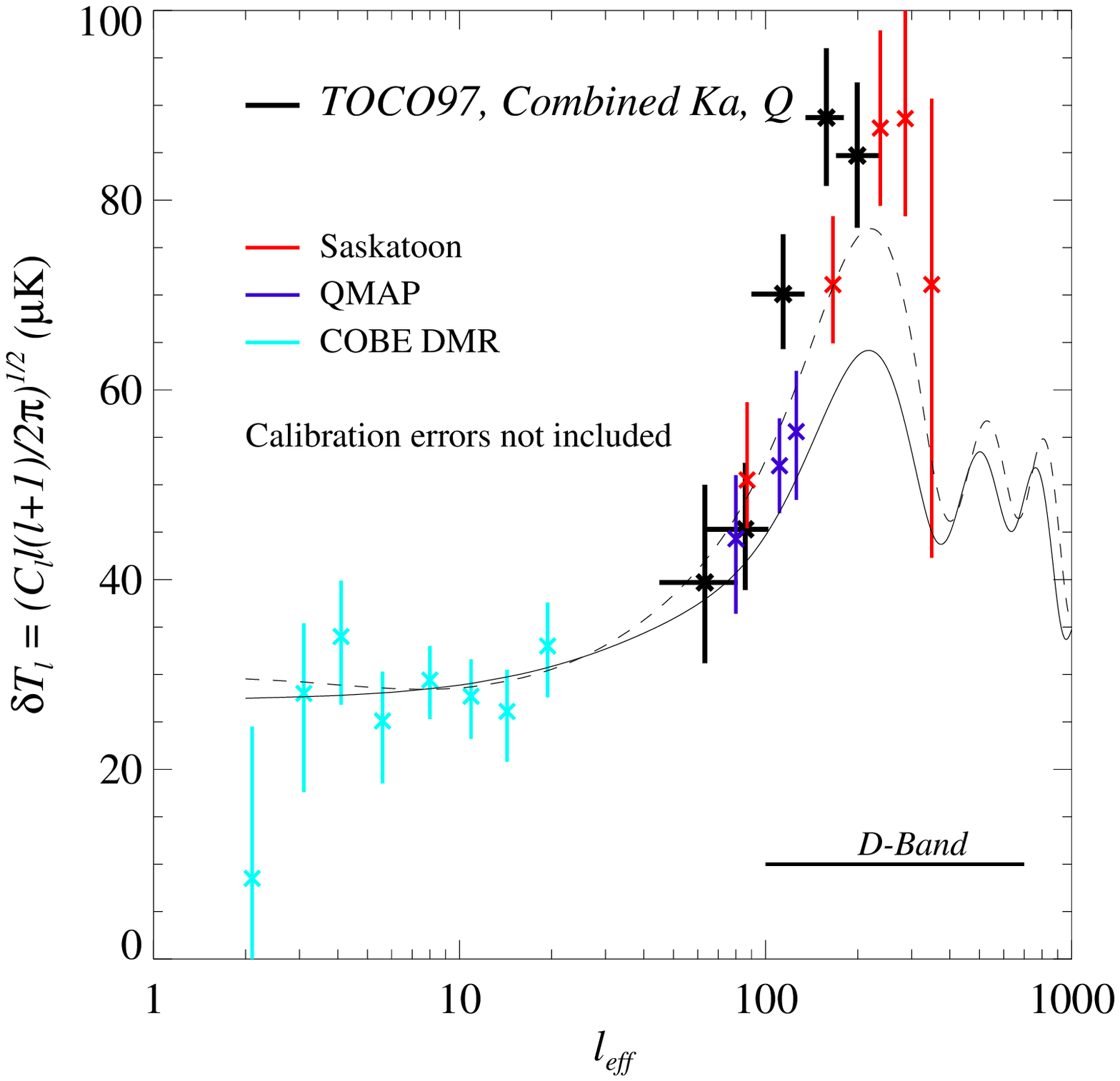}
\figurenum{3}
\caption{
Angular spectrum from the {\sl SK}, {\sl QMAP}, and {\sl
TOCO97} experiments. The {\sl SK} data
from Netterfield {\it et al.} (1997, ApJ, 474:47) and Wollack {\it et
al.} (1997, ApJ 476:440)
have been recalibrated according to 
the Mason {\it et al.} (1997, astro-ph/9903383), leading to an increase of 5\%, and reduced
according to the forground contribution in de Oliveira-Costa {\it et al.}
(1997, ApJL, 482:L17), leading to a reduction of 2\%. The foreground reduction was 
applied uniformly. Because the {\sl SK} data are primarily at 40 GHz, the Mason {\it et al.}
results must be extrapolated. In addition, uncertainty in the measured
passband, measurement uncertainty, and beam uncertainty, must be taken into
accounted. This leads to a $1\sigma$ calibration error of 11\%.
The {\sl QMAP} data are the same as those reported in de Oliveira-Costa
{\it et al.} (1998, ApJL, 509:L78) and have an average calibration error
of 12\%, the correction for foreground emission is $\approx$ 2\% (de Oliveira-Costa {\it et al.}
1999, {\it in prep}), though has not yet been precisely determined and so is not included.
Both {\sl SK} and {\sl QMAP} are calibrated with respect to Cas-A. The
{\sl TOCO97} data, which have a 
calibration error of 10\%, are calibrated with respect to
Jupiter. A foreground contribution, which is expected to be small, has not
been subtracted. When calibration errors are included, all three independent
experiments agree. Two cosmological models are shown for reference. The lower is ``standard
CDM'' ($\Omega_0 = 1$, $\Omega_{B} = 0.05$, $h=0.5$); the higher one
is the current ``concordance model'' (Bachall {\it et al.} 1999 {\it in
prep}, Turner 1999 astro-ph/9904051) with 
$\Omega_0 = 0.4$, $\Omega_{B} = 0.045$, $\Omega_\Lambda = 0.6$, and $h=0.65$.  
For {\sl COBE}/DMR we use the results from Tegmark.
}
\end{figure}


\begin{thebibliography}{}

\bibitem[Bond et al. 1991]{bond91} Bond, J. R., Efstathiou, G., Lubin,
P. M., \&\ Meinhold, P. R. 1991, \prl, 66, 2179

\bibitem[Bond 1996]{bond96} Bond, J. R. 1996 {\it Theory and
Observations of the Cosmic Microwave Background Radiation}, in ``Cosmology and
Large-Scale Structure,'' Les Houches Session LX, August 1993,
ed. R. Schaeffer, Elsevier Science Press

\bibitem[Bond et al. 1998]{bond98} Bond, J. R., Efstathiou, G.,
\&\ Tegmark, M. 1998, \mnras, 50, L33

\bibitem[Bond et al. 1998b]{bond98b} Bond, J. R., Jaffe, A. H., \&\ Knox,
L. 1998, \prd, 57, 2117

\bibitem[Coble et al. 1999]{cob99} Coble, K., et al. 1999, astro-ph/9902195

\bibitem[de Oliveira-Costa et al. 1997]{doc97} 
de Oliveira-Costa, A., Kogut, A., Devlin, M. J., Netterfield, C.
B., Page, L. A., \&\ Wollack, E. J. 1997 \apjl, 482, L17

\bibitem[Devlin et al. 1998]{dev98} Devlin, M. J., de
Oliveira-Costa, A., Herbig, T., Miller, A. D., Netterfield, C.
B., Page, L., \&\ Tegmark, M.  1998, \apjl, 509, L73

\bibitem[Griffin et al. 1986]{gri86} Griffin, M. J., Ade, A. R.,
Orton, G. S., Robson, E. I., Gear, W.K., Nolt, I. G., \&\
Radostitz, J. V. 1986, Icarus, 65, 244

\bibitem[Hu et al. 1997]{whu97} Hu, W., Sugiyama, N., \&\ Silk,
J. 1997, \nat, 386, 37

\bibitem[Jungman et al. 1995]{jung95} Jungman, G., Kamionkowski,
M., Kosowsky, A., \&\ Spergel, D. N. 1995, \prd, 54, 1332

\bibitem[Netterfield et al. 1997]{net97} Netterfield, C. B., Devlin, M. J.,
Jarosik, N., Page, L., \&\ Wollack, E. J. 1997, \apj, 474, 47

\bibitem[Kerr et al. 1993]{ker93} Kerr, A. R., Pan, S.-K.,
Lichtenberger, A. W., \&\ Lloyd, F. L.  1993, Proceedings of the 
Fourth International Symposium on Space Terahertz Technology, 1

\bibitem[Pospieszalski 1992]{posp92} Pospieszalski,
M. W. 1992, IEEE MTT-S Digest, 1369;
also see Pospieszalski, M. W. et al. 1994, IEEE MTT-S Digest, 1345

\bibitem[Pospieszalski et al. 1997]{posp97} Pospieszalski,
M.W. 1997, Microwave Background Anisotropies
(Ed. Frontieres, ed. Bouchet et al.), 23

\bibitem[Tegmark 1997]{max97} Tegmark, M., 1997,
\prd, 55, 5895

\bibitem[Ulich et al. 1981]{uli81} Ulich, B. L. 1981,
\aj, 86, 1619

\bibitem[Wollack et al. 1997]{wol97} Wollack, E. J., Devlin, M. J.,
Jarosik, N.J., Netterfield, C. B., Page, L., \&\ Wilkinson, D. 1997, \apj,
476, 440

\end{thebibliography}
\end{document}